\documentclass{elsart}

\usepackage{epsf}
\usepackage{amssymb}

\begin{document}

\begin{frontmatter}

\title{Classical behaviour of many-body systems in Bohmian Quantum Mechanics}

\author{H. Geiger\thanksref{emailharald}},
\author{G. Obermair\thanksref{emailgustav}} and
\author{Ch. Helm\thanksref{emailhelm}}

\address{ Institut  f\"ur Theoretische  Physik,  Universit\"at  Regensburg,
D-93040 Regensburg,  Germany }

\thanks[emailharald]{harald.geiger@physik.uni-regensburg.de}
\thanks[emailgustav]{gustav.obermair@physik.uni-regensburg.de}
\thanks[emailhelm]{christian.helm@physik.uni-regensburg.de}

\begin{abstract}
The classical behaviour of a macroscopic system consisting of a large number of
microscopic systems is derived in the framework of the Bohmian interpretation
of 
quantum mechanics. Under appropriate assumptions concerning the localization 
and factorization of the wavefunction it is shown explicitly that the center
of mass motion of the system is determined by the classical equations of
motion. 
\end{abstract}


\end{frontmatter}

\section{Introduction}
A deterministic formulation of quantum mechanics, which is equivalent to the 
usual theory with respect to the prediction of experimental results, 
had  originally been proposed by D. Bohm \cite{Bo52} and attracted some interest 
recently \cite{Al94,Bo93,Cu96,Ge98,Ho93}.
This alternative interpretation provides additional 
insight in various topics in quantum mechanics like the definition of 
quantum chaos analogously to the classical case \cite{Sch95,Du92c,De96} 
or the derivation of the statistical postulates of a measurement process 
from a deterministic theory \cite{wir}. 

Due to the nature of the theory containing  a classical particle, whose 
dynamics is determined by the quantum mechanical wavefunction, 
the transition from quantum behaviour for elementary particles and classical
motion of heavy point particles is continous. In this framework 
 the classical limit is simpler and more intuitive than in  the conventional
theory, avoiding e.g. the introduction of coherent states \cite{Bo93,Ho93,Bo85}. 

The problem, why normal macroscopic objects, which are not elementary, but 
contain a large number of microscopic, quantum mechanical subsystems, behave 
classically, is much less trivial and has up to now 
not been worked out in the framework of the Bohmian theory.
An additional important issue is to understand the relation between 
the nonpredictibility features of the quantum mechanical 
measurement process \cite{Bo93,Ge98,Ho93,Bo85} and a  classical 
measurement, which does not affect the state of the system. 
The proof that the center of mass motion of a macroscopic 
system is under certain assumptions
classical will be the main aim of the following investigations. 
The necessary characterizations of the wavefunction will thereby have a 
simple intuitive meaning in terms of forces on the Bohmian particle.

The paper is organized as follows: After a short review of the Bohmian 
formulation of quantum mechanics (chap. \ref{kapbohm}), the classical limit of 
macrocopic objects is characterized (chap. \ref{macrosection}) 
and the classical equations 
of motion for the center of mass coordinates are derived (chap. 
\ref{centerchap}).

\section{Bohmian quantum mechanics \label{kapbohm}}

Within the Bohmian mechanics \cite{Al94,Bo52,Bo93,Cu96,Ge98,Ho93}, a state 
of a system is
completely determined not only by the wavefunction $\Psi$, but also  
by the position $x(t)=(x_1,\ldots,x_N)(t)$ of 
a hidden particle in the configuration
space of the system.

The dynamics of the wavefunction $\Psi$ is determined by the
Schr\"odinger equation in the usual way, while
the dynamics of the particle is deduced from the wavefunction
$\Psi(x,t)$.

By introducing the modulus $R(x,t)$ and the phase $S(x,t)$ of the
wavefunction 
$ \Psi(x,t)=R(x,t) e^{\frac i \hbar S(x,t)}$
the Schr\"odinger equation  can be written as 
\begin{equation} -\frac{\partial }{\partial t}S(x,t) =\frac { (\frac \partial {\partial
x} S(x,t))^2 }{2m} - \frac{\hbar^2}{2m} \frac{\frac{\partial^2}{\partial
x^2} R(x,t)}{R(x,t)} + V(x,t), \label{HJ}
\end{equation}

\begin{equation} \frac{\partial }{\partial t}R(x,t)^2 +\frac \partial {\partial
x}\Big(R(x,t)^2 \frac{\frac{\partial}{\partial
x}S(x,t)}{m}\Big)  = 0. \label{Ko}
\end{equation}

While equation (\ref{Ko}) represents a continuity equation of 
$\vert\Psi(x,t)\vert^2$,
equation (\ref{HJ}) is a Hamilton Jacobi equation of a classical particle
with coordinate $x$ in the presence of both the classical potential
$V(x,t) $  and the so-called quantum potential \cite{Bo52} 
\begin{equation}
Q(x,t)  := - \frac{\hbar^2}{2m} \frac{\frac{\partial^2}{\partial
x^2} R(x,t)}{R(x,t)} \quad . \label{quantum_potential}
\end{equation}

In particular the 
momentum $p(t)$ and the energy $E(t)$  of the particle are given by 
derivatives  of the phase $S(x,t)$ of the wavefunction $\Psi(x,t)$:
\begin{equation} 
p(t):= m { \dot x} =  \frac\partial{\partial x} S(x,t)\vert_{x(t)}
\label{pH}
\quad , \quad 
 E(t):=-\frac\partial{\partial t} S(x,t)\vert_{x(t)} \quad . 
\end{equation}
This first-order differential equation is sufficient to calculate the 
trajectory $x (t)$ from the initial value $x(t_0)$, but the 
relative significance of classical and quantum mechanical effects becomes
clearer in a different representation.
After a straightforward calculation (cf. appendix \ref{appendix1})   
the equation of motion of the Bohmian particle $x(t)$
can be presented in a form, which is similar to Newton's law, but contains 
a quantum force $F_Q := - \frac{\partial}{\partial x} Q$ in addition:
\begin{equation} 
\frac d {dt} p(t)=-\frac\partial{\partial x}\Big(
V(x,t)+Q(x,t)\Big)\Big\vert_{x(t)} = : F (t) \quad .  \label{2F}
\end{equation}

\section{Characterisation of classical many-body systems
\label{6.1} \label{macrosection}}

One of the intuitive features of the Bohmian formulation of quantum 
mechanics is the fact that there is a continous and simple limit from the 
quantum to the  classical regime of a single degree of freedom. 
By increasing the mass $m$ of the particle, 
i.e. $\frac{\hbar^2}{2m} \rightarrow 0$, the quantum force 
$F_Q = - \partial_x Q \rightarrow 0$ in equ. \ref{2F} vanishes and 
the quantum potential $Q = {\rm const.}$ becomes irrelevant in the 
Hamilton-Jacobi equation \ref{HJ}. The absence of quantum mechanical
 interactions $\sim Q$ in addition to 
 the classical potential $V(x)$ also provides a natural 
explanation, why the state of a classical system of this kind will not be 
affected during  a measurement.

If and under what circumstances the collective motion of objects 
consisting out of 
a macrocopic number $N \sim 10^{23}$ of quantum mechanical systems of mass $m$
with $\frac{\hbar^2}{2m} \sim O(1)$ is classical, is far less obvious and 
will be investigated in the following. 
Thereby it turns out that additional assumptions 
have to be made in order to recover classical physics and to
exclude macroscopic quantum phenomena like superconductivity or superfluidity.

Let the macrocopic system consist of $N$ subsystems, e.g. atoms, with 
center of mass coordinates $x_i(t)$ for  $i\in\{1,\ldots ,N\}$, which
are determined by the wavefunction 
\begin{eqnarray}
\Psi_{\rm sym} (x_1,\ldots ,x_N,t)
&=& \sum_{\nu=1}^{N!}
c_\nu\Psi (x_{\pi_\nu(1)},\ldots ,  x_{\pi_\nu(N)},t) = \label{6neu} \\
&=:& \sum_{\nu=1}^{N!}
c_\nu \Psi_{\nu} (x_1,\ldots ,  x_N ,t) 
\quad . 
\end{eqnarray}
Thereby the exchange symmetry of the indistinguishable particles has been 
built in by a summation over the permutations $\pi_\nu$ 
($\nu=1,\dots,N! $) with appropriate coefficients $c_\nu$.

Having in mind classical systems like solids, fluids or gases consisting of 
a large number of atoms, which show no quantum behaviour 
like Bose-Einstein condensation, we can assume the following  two
features:

{\rm \em (i) Locality:} 
  The quantum mechanical motion of the center of mass 
  coordinates of  different subsystems,
  which are determined by the wavefunction $|\Psi|^2$  via equ. \ref{HJ}, 
  are assumed to be confined to disjunct regions in configuration space. 
  This can be motivated by the physical picture that the quantum mechanical 
  uncertainity of the position of the center of mass is much smaller than 
  the minimal distance, i.e. the radius of the atom, of the subsystems.

More explicitly this feature manifests itself as a vanishing overlap of
the functions $\Psi_\nu$ corresponding to different permutations of the 
center of mass coordinates: 
\begin{equation}
\int\!\! dx_1\cdots d x_N\,\Psi^*(x_{\pi_\nu(1)},\ldots
x_{\pi_\nu(N)},t)\Psi(x_{\pi_{\tilde\nu}(1)},\ldots
x_{\pi_{\tilde\nu}(N)},t)= \delta_{\nu {\tilde \nu}} \; . \label{6p0} 
\end{equation}
In the Bohmian interpretation this means that the particle
 $(x_1,\dots,x_N)(t)$ cannot tunnel from the support 
$\Gamma_\nu := \{ x \in \Rset^N \vert \Psi_{\nu} (x) = \partial_i \Psi_{\nu} 
(x) = 0 \quad \forall i \} $ 
of one wavepacket $\Psi_{\nu}$ to another. 
Consequently the ergodicity of its trajectory $x(t)$ is restricted to a certain
subset $\Gamma_{\mu}$ and the other parts $\Psi_\nu$ ($\nu \neq \mu$) 
of the wavefunction can be neglected without loss of generality. 
Note that here the selection of a certain sector $\mu$ is given by the 
position of the particle and not due to the collapse of the wavefunction 
after a measurement. 

\begin{figure}
\leavevmode
\begin{center}
\epsfxsize=\textwidth
\epsfbox{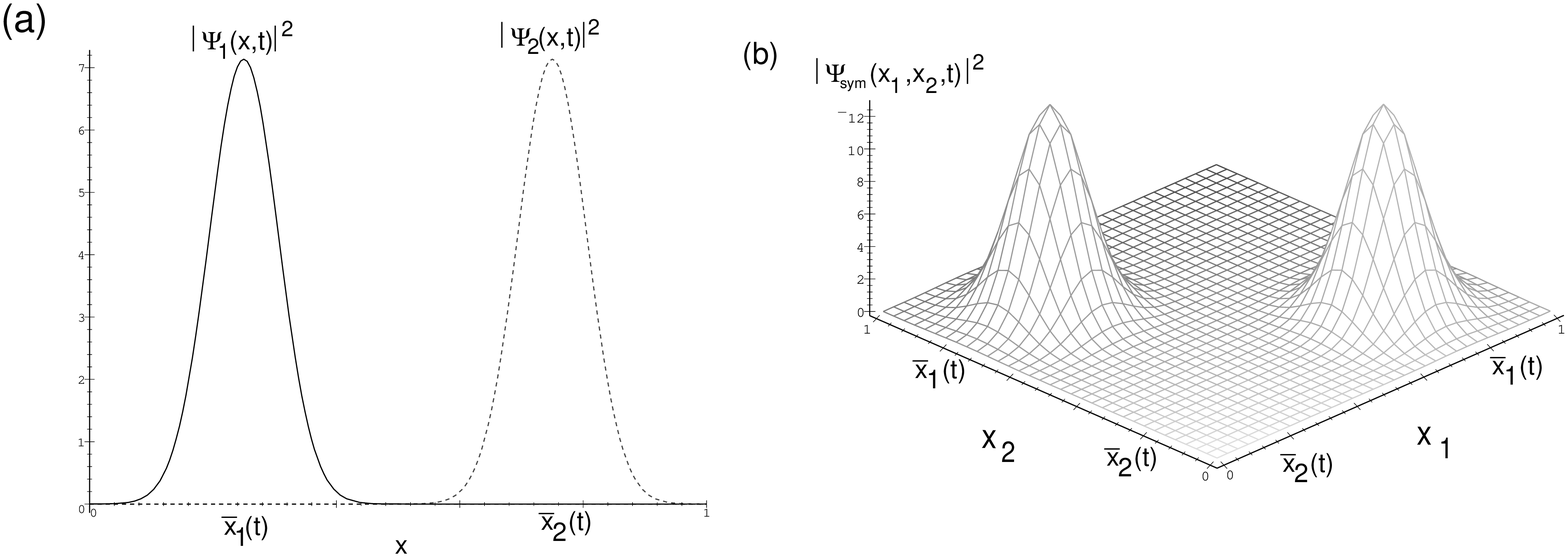}
\end{center}
\caption{\label{Abb.6.2} 
Schematic wavefunction of a system consisting of $N=2$
 one\-di\-men\-sio\-nal subsystems:
(a) The probability distributions  $\vert\Psi_i(x,t)\vert^2$ ($i=1,2$) 
are localized around the values $\bar{x}_{i}$ and have vanishing overlap 
according to assumption (i) (cf. equ. \ref{6p0}).
(b) The modulus  $\vert\Psi_{\rm sym}(x_1,x_2,t)\vert=\frac{1}{\sqrt{2}}
\Big(\vert\Psi_1(x_1,t)\vert\,\vert\Psi_2(x_2,t)\vert + 
\vert\Psi_2(x_1,t)\vert\, \vert\Psi_1(x_2,t)\vert\Big)$ of the total 
symmetrized wave function is localized around maxima at 
$(\bar{x}_1,\bar{x}_2)$ (i.e. $\nu=1$)   and $(\bar{x}_2,\bar{x}_1)$
(i.e. $\nu=2$).
Due to the vanishing overlap no transition of the system particle
$(x_1,x_2)$ between the support $\Gamma_{\nu}$ of the two wavepackets
$\nu=1,2$ is possible and it is sufficient to keep the part $\Gamma_\mu$,
where the particle is located. 
}
\end{figure}

As an example the expectation value 
\begin{equation} 
\bar{x}_i (t):= \int\!\! dx_1 \cdots \int\!\! dx_N \,
\Psi^{*}_{\mu}(x_1,\dots,x_N,t) x_i\Psi_{\mu}(x_1 ,\ldots,x_N,t)
\end{equation}
of the center of mass coordinate $x_i$ has to be averaged over the support
$\Gamma_\mu$ 
only. Note that $\bar{x}_i(t)$ denotes the average position of the center of mass
in an infinite number 
of measurements and is not identical with the position $x_i(t)$ of the 
Bohmian particle. The locality assumption (i) can therefore be expressed more 
precisely by the statement that 
$\vert x_i(t)-\bar{x}_{ i}(t)\vert$ is small compared with the minimal
 distance of the center of mass coordinates of the subsystems. 
The wavefunction $\Psi_{\rm sym} (x_1,x_2,t)$ is expressed schematically
for the 
two-dimensional case in figure \ref{Abb.6.2} and  the index $\mu$ of 
the wavefunction $\Psi_\mu$ will be suppressed in the following. 

{\rm \em (ii) Factorization:}
  In addition to this, the factorization of the wavefunction
  \begin{equation}
    \Psi(x_1,\ldots ,x_N,t)=R_1(x_1,t)\cdots
    R_N(x_N,t)e^{\frac i \hbar S(x_1,\ldots,x_N,t)} \label{factorization}  
  \end{equation}
  within each disjunct region $\Gamma_\nu$
  turns out to be essential for the elimination of macroscopic 
  quantum mechanical correlations. This assumption corresponds to the 
(ad hoc)  introduction of "decoherence" for the reconstruction of classical
 mechanics from the conventional quantum theory
  \cite{Ge98} and will be justified intuitively in the following.

As a consequence, the modulus
\begin{equation}
\vert\Psi (x_1,\ldots,x_N,t)\vert
=\vert\Psi_1(x_1,t)\vert\cdots\vert\Psi_i(x_i,t)\vert\cdots
\vert\Psi_N(x_N,t)\vert \label{factorization2} 
 \end{equation}
factorizes due to the  decoherence assumption (ii) of equ.
\ref{factorization}
and the quantum potential 
\begin{equation}
Q(x_1,\ldots,x_N,t)=Q_1(x_1,t)+\cdots +Q_N(x_N,t) \label{qqq}
\end{equation}
is a sum of one particle potentials. This means that the motion 
\begin{equation} 
m_i\ddot x_i=-\frac\partial {\partial x_i}
V(x_1,\ldots,x_N,t)\Big\vert_{x(t)}-\frac\partial {\partial x_i}
Q_i(x_i,t)\Big\vert_{x_i(t)} 
\end{equation}
of the center of mass of a subsystem is correlated to the other subsystems 
via the classical potential $V(x_1,\dots,x_N)$ only. 
Here the physical meaning of equ. \ref{factorization} becomes  transparent:
Short-range quantum mechanical processes within the subsystem $i$ can affect 
its center of mass motion,  but have no effect on the other subsystems.

Formula equ. \ref{factorization2} can be simplified by noting that the 
probability distributions $\vert \Psi^{l(i)} (x,t) \vert^2$ of the Bohmian 
particles $x_i$
around the mean value $\bar{x}_i$ are identical for equivalent subsystems 
of the same type $l(i) \in \{ 1, \dots, n  \}$:
\begin{equation}
\vert\Psi (x_1,\ldots,x_N,t)\vert =
\prod_{i=1}^{N}\vert\Psi^{l(i)}\big(x_i-\bar{x}_{i}(t),t \big) \vert \quad .
 \label{6xyz}\end{equation}

The condition $N \gg n$ that there is a macroscopic number of subsystems of
each type is fulfilled for macroscopic objects like solids 
with a small number of inequivalent sites or gaseous mixtures of different
particles and is at least approximately true for amorphous glass-like objects.

\section{Center of mass motion \label{6.1.b} \label{centerchap}}

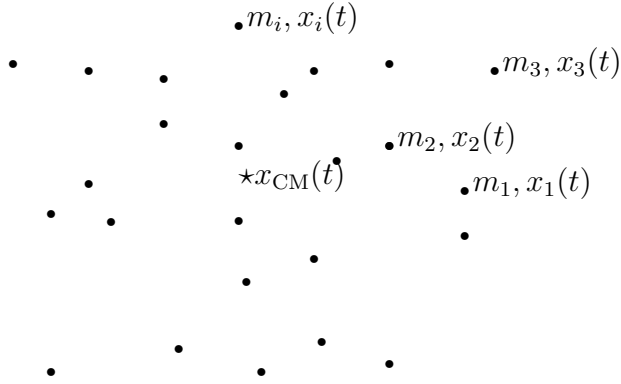
\begin{figure}
\setlength{\unitlength}{1cm}
\begin{picture}(10,5)
\put(5,2.5){$\star x_{\rm CM} (t)$}
\put(8,2.4){\circle*{0.1}$m_1,x_1(t)$}
\put(7,3){\circle*{0.1}$m_2,x_2(t)$}
\put(8.4,4){\circle*{0.1}$m_3,x_3(t)$}
\put(5,4.6){\circle*{0.1}$m_i,x_i(t)$}
\put(8,1.8){\circle*{0.1}}
\put(7,3){\circle*{0.1}}
\put(6.3,2.8){\circle*{0.1}}
\put(5.6,3.7){\circle*{0.1}}
\put(3,4){\circle*{0.1}}
\put(2,4.1){\circle*{0.1}}
\put(3.3,2){\circle*{0.1}}
\put(6,4){\circle*{0.1}}
\put(4,3.9){\circle*{0.1}}
\put(3,2.5){\circle*{0.1}}
\put(7,4.1){\circle*{0.1}}
\put(2.5,2.1){\circle*{0.1}}
\put(5,2){\circle*{0.1}}
\put(6,1.5){\circle*{0.1}}
\put(4,3.3){\circle*{0.1}}
\put(5,3){\circle*{0.1}}
\put(7,0.1){\circle*{0.1}}
\put(2.5,0.0){\circle*{0.1}}
\put(5.1,1.2){\circle*{0.1}}
\put(6.1,0.4){\circle*{0.1}}
\put(4.2,0.3){\circle*{0.1}}
\put(5.3,0.0){\circle*{0.1}}
\end{picture}

\bigskip

\caption{\label{Abb.6.3} 
$N$ microscopic quantum mechanical 
systems with the center of mass coordinates $x_i(t)$
and masses $m_i$  ($i\in\{1,\ldots,N\}$) form a macroscopic object with 
center of mass $x_{\rm CM} (t)$ (at the point $\star$), his dynamics being 
governed by classical equations of motion.
 }
\end{figure}

The center of mass coordinate 
\begin{equation}
x_{\rm CM} (t):=\frac 1 {M} \sum_{i=1}^N x_i(t) m_i
\end{equation}
of a macroscopic object with mass $M$ is given by the average of the center of
mass coordinates $x_i$ of the microscopic subsystems.

Although the trajectory $x_{\rm CM} (t)$ can be derived from equ. 
\ref{pH} alone, this is not convenient, as the phase $S(x_1, \dots, x_N,t)$
depends not only on $x_{\rm CM}$, but on all variables $x_i$ and must be
calculated by a solution of the full Schr{\"o}dinger equation. 

Using equ. \ref{2F} for the quantum mechanical dynamics of $x_i (t)$ 
instead one obtains
\begin{equation}
  \ddot{x}_{\rm CM}(t) M=-\sum_{i=1}^N \frac \partial {\partial
x_i}V(x_1,\dots,x_N,t)\Big\vert_{x (t)} 
\! \!- \!\! \sum_{i=1}^N \frac \partial {\partial
x_i}Q(x_1,\dots,x_N,t)\Big\vert_{x (t)}  . \label{6VQ}
\end{equation}
In the classical contribution
\begin{equation} 
-\sum_{i=1}^N\frac \partial {\partial
x_i}V(x_1,\ldots,x_N,t)\Big\vert_{x(t)} = 
\sum^N_{{i,j} \atop {i \neq j}} F_{ij} +
\sum_{i=1}^N F_i^{ext}=:F^{\rm ext}
\label{6V}
\end{equation}
the internal forces $F_{ij}$ from subsystem $j$ on $i$ cancel due to 
Newton's law $F_{ij} = - F_{ji}$ and only the total external force 
$F^{\rm ext}$ remains.
Due to equ. \ref{6xyz} the quantum force for a homogeneous 
system ($n=1$) is given by
\begin{equation} 
F_Q^{\rm CM} := 
- \sum_{i=1}^N \frac \partial {\partial
x_i}Q(x_1,\dots,x_N,t)\Big\vert_{x(t)} \! 
=-\sum_{i=1}^N\frac \partial {\partial
x_i} Q\big(x_i-\bar{x}_i(t),t\big)\vert_{x_i(t)}  , \label{6Q}   
\end{equation}
where 
\begin{equation}  
Q\big( x_i-\bar{x}_{i}(t),t \big) := -\frac{\hbar^2}{2m}\frac{\frac{\partial^2}
{\partial
x_i^2}\vert\Psi
\big(x_i-\bar{x}_{i}(t),t\big)
\vert}{\vert\Psi\big(x_i-\bar{x}_{i}(t),t\big)\vert} \; \; .
\end{equation}
For large $N \gg 1$ it is possible to replace the sum in equ. 
\ref{6Q} by an integral. Thereby it is necessary to introduce the distribution 
function ${\mathcal{P}}(x_i-\bar{x}_{i}(t),t)$ of the position $x_i(t)$ of 
the system particle around the center of mass $\bar{x}_i$. This is given 
by the condition  ${\mathcal{P}}(x_i-\bar{x}_{i}(t),t)=\vert\Psi
\big(x_i-\bar{x}_{i}(t),t\big)\vert^2$ of so-called 
quantum equilibrium \cite{Du92a}, which can be derived from the quasi-ergodic 
dynamics of the system particle during a sequence of measurements
\cite{Ge98,wir}.
This leads to 
\begin{eqnarray}
F_Q^{\rm CM} &:=& -\frac 1 N \sum_{i=1}^N\frac \partial {\partial
x_i}Q\big(x_i-\bar{x}_{i}(t),t\big)
\Big\vert_{x_i(t)}\approx \\
& \approx & 
-\int_{-\infty}^\infty\! du\,
\vert\Psi\big(u-\bar{x}_{i}(t),t\big)\vert^2 \frac \partial {\partial
u}Q\big(u-\bar{x}_{i}(t),t\big) \label{6sumint} = 0 \; , 
\end{eqnarray}
which is shown to vanish in appendix \ref{appendix2}. Thereby the error can be 
estimated to be smaller than $\frac{\vert F_{Q_{\rm max}} \vert}{N}$, 
$ F_{Q_{\rm max}}$ being the largest single quantum force. Hence the total 
quantum force 
\begin{equation} 
\Big\vert\sum_{i=1}^N \frac\partial{\partial x_i}
Q(x_1,\ldots,x_N,t)\Big\vert_{x_1(t),\ldots,x_N(t)}\leq\vert F_{Qmax}\vert. \label{6Qabs} 
\end{equation} 
on the center of mass is of microscopic magnitude and can therefore not 
accelerate the macroscopic object of mass $M$ substantially. This
proofs  Newton's Law
\begin{equation} 
\ddot{x}_{\rm CM} (t)M= F^{ext} \quad  \label{6erg}
\end{equation}
for the collective center of mass motion.
This result is also true in the general case $n\neq 1$, as  in the case 
$n \ll N$ the above argument can  be used independently for each type
$l(i) \in 1, \dots, n$ of subsystems.

An important corollary of the calculation presented here is that there is no
quantum mechanical interaction between collective classical coordinates 
of two macroscopic objects, one representing a measurement device and the 
other one being the system to be measured. Therefore in a classical 
measurement the  effect of the quantum potentials of the underlying 
microscopic subsystems neither causes any uncertainity in the measurement
result nor affects the system to be measured beyond the classical interaction.

In order to clarify our characterisation of conventional classical systems 
further, we would like to discuss briefly the case of a 
Bose-Einstein condensate
\begin{equation} 
\Psi(x_1,\ldots,x_N,t)=R(x_1,\dots, x_N,t)e^{\frac i \hbar
\Big(S(x_1,t)+\cdots +S(x_N,t)\Big)}  
 \sim e^{\frac i \hbar \sum\limits_{i=1}^Nm v x_i}
\label{6sw} \!  ,
\end{equation}
where coherence of the phases $S_i(x_i,t) = S (x_i,t)$ of the subsystems is
essential. 
Here the motion of the center of mass coordinate can 
most easily determined from equ. \ref{pH} directly as
\begin{equation}
 \dot{x}_{CM} (t)=v  \Longrightarrow 
x_{CM} (t) = v ( t - t_0  ) + x_{CM} ( t_0 ) \quad . 
\end{equation}
Here the fundamental nature of a macroscopic quantum phenomen in contrast to
conventional classical objects becomes evident: As the acceleration
$\ddot{x}_{CM}=0$ vanishes, the total quantum force is of the same order as
the classical one, while for the system obeying equ. \ref{factorization2}
 the effect of quantum forces averages out on a macroscopic level.

\section{Summary and Conclusions}

The reconstruction of classical dynamics for ordinary macroscopic objects 
containing subsystems, which behave quantum mechanically, has been reviewed 
from the viewpoint of Bohmian quantum mechanics. 
In this formulation the quantum mechanical effects are contained exclusively
in an additional quantum force appearing in the classical equations of motion 
of the system particle. 
In the case of a point particle of macroscopic mass 
($\hbar^2 / 2 m \rightarrow 0$) the quantum potential 
becomes irrelevant, defining the classical limit in a conceptually clear way. 
In contrast to this, it was shown that additional restrictions on a 
many-body wavefunction have to be employed to recover Newton's Law for the 
center of mass motion. 

One main assumption concerned the fact that the wave function is 
sufficiently localized to prevent the Bohmian center of mass particle 
of an atom from leaving this subsystem. In addition to this, an appropriate 
factorization of the modulus $\vert \Psi  \vert$ of the wavefunction is 
essential (cf. equ. \ref{factorization}) in order to eliminate quantum
mechanical
correlations between the different (quantum mechanical)  subsystems. 
In this case  quantum phenomena can be ruled out on macroscopic scales and 
the effect of quantum forces on collective macroscopic variables 
averages out. 
The Bohmian interpretation thereby provides further insight 
into the nature of the classical limit by suggesting
an intuitive physical picture and motivation for the essential features 
of the  wavefunction.

\begin{appendix}

\section{The force on the Bohmian particle \label{appendix1}}

The force $F(t)$ on the Bohmian particle is given by
\begin{eqnarray} 
F(t)  &:=& \frac d {dt} p(t) = \frac d {dt}\frac\partial {\partial x}
S(x,t)\vert_{x(t)} =  \\
&=& \Big( \frac\partial{\partial x} \frac\partial {\partial t}
S(x,t) + \frac{\partial^2}{\partial x^2} S(x,t) \frac{d x}{d t}\Big)
\Big\vert_{x(t)} \quad .
\end{eqnarray}
Using equation \ref{pH} one obtains
\begin{eqnarray} 
F(t)&=&\Big(\frac\partial{\partial x} \frac\partial {\partial t}
S(x,t) +\frac 1 m \frac{\partial^2}{\partial x^2} S(x,t)
\frac\partial{\partial x}S(x,t)\Big)\Big\vert_{x(t)} = \\
&=&
 \frac\partial{\partial x}\Big(\frac \partial {\partial t}
S(x,t)+\frac 1 {2m} \Big(\frac\partial{\partial x}
S(x,t)\Big)^2\Big)\Big\vert_{x(t)} \quad .
\end{eqnarray}
With the help of the Hamilton-Jacobi equation
\begin{equation}
\frac\partial {\partial t}
S(x,t)+\frac 1 {2m}\Big(\frac\partial{\partial x}
S(x,t)\Big)^2+V(x,t)+Q(x,t)=0  \label{hamilton_jacobi}
\end{equation}
equation \ref{2F} is proven:
\begin{equation} 
F(t)=-\frac \partial {\partial
x}\Big(V(x,t)+Q(x,t)\Big)\vert_{x(t)}  \quad . 
\end{equation}

\section{Averaging of quantum forces}

\label{appendix2}

It will be shown that
\begin{equation}
I :=  \int_{-\infty}^\infty\! dx\,\vert\Psi(x,t)\vert^2
\frac\partial{\partial x}Q(x,t)=0 \label{A C} \quad ,
\end{equation}
 $x \in \Rset^N$ being a multidimensional vector and 
$\vert\Psi(x,t)\vert=:R(x,t)$ a square-integrable function with 
$R(x) \rightarrow 0$ and $\partial_x R(x) \rightarrow 0$ for 
$\vert x \vert
 \rightarrow \infty$ and the quantum potential 
$Q(x)$ as defined in equ. \ref{quantum_potential}. 
One easily obtains
\begin{eqnarray} 
I &=& -\frac{\hbar^2}{2m}\int_{-\infty}^\infty\! dx\,R^2(x,t)\frac {\partial}
{\partial x}\Big(\frac{\frac{\partial^2}{\partial
x^2} R(x,t)}{R(x,t)}\Big)  = \\
&=&  -\frac{\hbar^2}{2m}\int_{-\infty}^\infty\! dx\,
\Big(R(x,t)\frac{\partial^3}{\partial
x^3} R(x,t)-\frac{\partial^2}{\partial
x^2}   R(x,t)\frac{\partial}{\partial
x} R(x,t)\Big) = \\
&=&  \frac{\hbar^2}{m}\int_{-\infty}^\infty\! dx\,
\frac{\partial^2}{\partial
x^2}   R(x,t)\frac{\partial}{\partial
x} R(x,t)  \quad ,
\end{eqnarray}
where in the last step a partial integration has been performed. An additional
partial integration shows that $R \equiv - R =0$ proofing equ. \ref{A C}.

\end{appendix}

\end{document}